\documentclass[aps,prl,twocolumn,groupedaddress]{revtex4-1}
\usepackage{graphicx}
\usepackage{amsfonts}
\usepackage{amsmath}
\usepackage{color}
\usepackage{soul}
\newcommand{\ket}[1]{\vert#1\rangle}
\newcommand{\bra}[1]{\langle#1\vert}

\def\opone{\leavevmode\hbox{\small1\kern-3.8pt\normalsize1}}

\begin{document}

\title{Entanglement swapping with quantum-memory-compatible photons}

\author{J. Jin$^1$} \altaffiliation{Present address: Institute for Quantum Computing, Department of Physics and Astronomy, University of Waterloo, 200 University Ave West, Waterloo, Ontario, Canada}
\author{M. Grimau Puigibert$^1$}
\author{L. Giner$^1$}\altaffiliation{Present address: Department of Physics, University of Ottawa, 150 Louis Pasteur, Ottawa, Ontario, Canada}
\author{J. A. Slater $^{1}$}\altaffiliation{Present address: Vienna Center for Quantum Science and Technology, Faculty of Physics, University of Vienna, Boltzmanngasse 5, A-1090, Vienna, Austria}
\author{M. R. E. Lamont$^1$}
\author{V. B. Verma$^2$} 
\author{M. D. Shaw$^3$} 
\author{F. Marsili$^3$} 
\author{S. W. Nam$^2$} 
\author{D. Oblak$^1$}
\author{W. Tittel$^1$}\email{wtittel@ucalgary.ca}
\affiliation{$^1$ Institute for Quantum Science and Technology, Department of Physics $\&$ Astronomy,
University of Calgary, 2500 University Drive NW, Calgary, Alberta T2N 1N4,Canada, $^2$ National Institute of Standards and Technology, Boulder, Colorado 80305
USA, $^3$ Jet Propulsion Laboratory, California Institute of Technology, 4800 Oak Grove Drive, Pasadena, California 91109, USA}

\date{\today}

\begin{abstract}
We report  entanglement swapping  with time-bin entangled photon pairs, each constituted of a 795~nm photon and a 1533~nm photon, that are created via spontaneous parametric down conversion in a non-linear crystal.  After projecting the two 1533~nm photons onto a Bell state, entanglement between the two 795~nm photons is verified by means of quantum state tomography. As an important feature, the wavelength and bandwidth of the 795~nm photons is compatible with Tm:LiNbO$_3$-based quantum memories, making our experiment an important step towards the realization of a quantum repeater.
\end{abstract}

\pacs{}

\maketitle

\section{Introduction}

Entanglement swapping entangles two photons that have no common past \cite{M. Zukowski 1993ab}. This fascinating phenomenon not only stimulated curiosity to understand quantum correlations \cite{E. Megidish 2013ab, X. -S. Ma 2012ab, A. J. Short 2006ab, C. Branciard 2012ab, W. Kłobus 2012ab, S. Bose 1998ab}, but also plays an important role in various applications of quantum information science, including quantum computing \cite{Ladd_10a, E. Knill 2001ab} and quantum repeaters \cite{H. -J. Briegel 1998ab}. A quantum repeater-based communication channel, for instance, exploits entanglement swapping to entangle interim nodes in a heralded fashion, and connect elementary entangled links (connecting interim nodes) to distribute entanglement in principle over arbitrarily long distances \cite{N. Sangouard 2011abc, N. Sinclair 2014abc}. In turn, the resulting entanglement can be used to generate a secret key  between distant users \cite{A. K. Ekert 1991ab}. 

Necessary ingredients for quantum repeaters, in addition to entangled photon pairs and entanglement swapping, are optical quantum memories \cite{Lvovsky_09a}. Such memories allow the reversible mapping of (entangled) states between light and atoms, and thereby remove the necessity for all elementary links to establish entanglement simultaneously. While entanglement swapping has been reported before \cite{J. W. Pan 1998abc, T. Jennewein 2002abc,  H. de Riedmatten 2005abc, M. Halder 2007a, Z. -S. Yuan 2008ab, R. Kaltenbaek 2009a}, the spectra of the resulting entangled photons were either orders of magnitude too large, or their wavelength were not suitable to allow subsequent interfacing with optical quantum memory. Here we remove this impediment: using cavities, we spectrally engineer swapped photons in time-bin qubit states in such a way that their wavelengths (around 795~nm) and bandwidths become compatible with our solid-state quantum memories \cite{E. Saglamyurek 2011ab}. Furthermore, we accordingly increase the coherence times of the photons used to swap the entanglement (both around 1533~nm wavelength), which will, in the future, allow such quantum interference measurements even with photons that have traveled through tens of kilometres of deployed standard telecommunication fibre~\cite{A. Rubenok 2013ab}. To verify successful entanglement swapping, we measure (conditional) two-photon visibility curves. For more complete information, we additionally employ quantum state tomography to derive the concurrence and the fidelity of the swapped state with the nearest maximally entangled state, and we compare the latter with the predictions of a recently developed model \cite{S. Guha 2014a}. As with the use of quantum state tomography, this has not been done previously for time-bin qubits after entanglement swapping.

\section{Experiment}

\begin{figure}[htpb]
\includegraphics[width=\columnwidth]{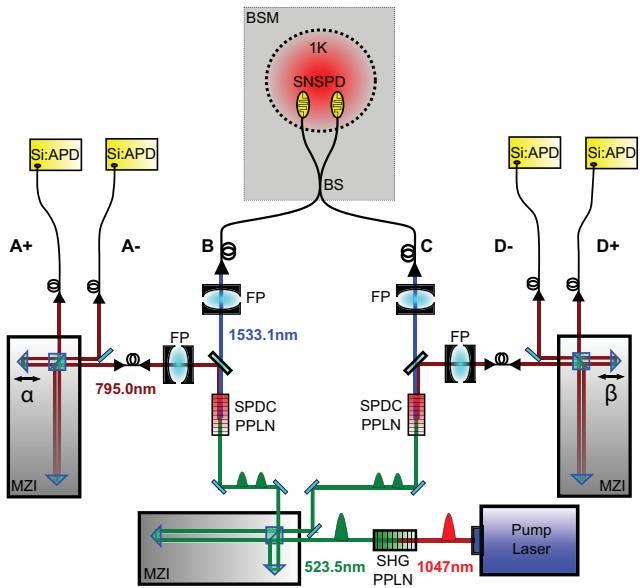}
\caption[(color online) Schematics of of our setup.]{Schematics of of our setup. See text for details.}
\label{entsetup}
\end{figure}

A schematics of our experimental setup is depicted in Fig. \ref{entsetup}. A 1047~nm wavelength laser emits 6 ps long pulses at 80 MHz repetition rate. After second harmonic generation (SHG) in a periodically poled lithium niobate crystal (PPLN), the now 18 ps long pulses, centered at 523.5~nm wavelength, travel through an unbalanced Mach-Zehnder interferometer (MZI) whose path length difference corresponds to 1.4 ns travel time difference, thereby splitting every pulse into two. Pairs of pulses emitted from the two outputs of the interferometer then pump two 10 mm-long PPLN crystals, in which spontaneous parametric down-conversion (SPDC) leads to time-bin entangled qubits  \cite{W.Tittel 2001ab} encoded into pairs of photons with wavelengths centred around 795~nm and 1533~nm. Assuming, for the sake of explanation, for the moment that only individual photon pairs are created, this process yields states of the form $\ket{\Phi^+}_{AB} = \frac{1}{\sqrt{2}}\left( \ket{e}_{A}\ket{e}_{B} + \ket{\ell}_{A}\ket{\ell}_{B}\right)$ emitted from one crystal, and $\ket{\Phi^-}_{CD} =\frac{1}{\sqrt{2}}\left( \ket{e}_{C}\ket{e}_{D} - \ket{\ell}_{C}\ket{\ell}_{D}\right)$ emitted from the other crystal, where $\ket{e}$ and $\ket{\ell}$ represent \emph{early} and \emph{late} time-bin qubit states.  

Using a grating monochromator connected to a single photon detector, we measure the spectral width of the 795~nm photon at full width half maximum (FWHM) to be 1.5~nm. By energy conservation, and taking into account the bandwidth of the pump photons, we calculate the spectral width of the 1533~nm photons to be 5.6~nm (FWHM). These values by far exceed the maximum bandwidth of 10 GHz (corresponding to 21 pm at 795~nm wavelength), over which quantum memories have so far been reported to operate \cite{E. Saglamyurek 2011ab}. Hence, to allow future interfacing with memories, we filter the 795~nm photons using Farbry-Perot cavities (FP; one per source) to 6 GHz, and also reduce the bandwidth of the 1533~nm photons to 12 GHz (again using Fabry-Perot cavities), which corresponds to 94 pm. (The filter cavities feature only one spectral order within the spectral width of the created photons.) This additionally ensures that the photons' coherence time, around 37 ps, exceeds that of the pump pulses, as required for entanglement swapping. The cavities are based on Corning ultra-low expansion glass spacers and fused silica mirrors to reduce thermal noise, and their transmissions are 35\% and 90\% for 795~nm and 1533~nm photons, respectively. Using the approach described in \cite{Marcikic2002}, we find the probability of having a photon pair per source and pump laser pulse after filtering, i.e. per qubit, to be about $19.1\pm1.8$\%.  

To swap entanglement to the two 795~nm qubits (propagating along spatial modes A and D), the two 1533~nm qubits (in modes B and C) are subjected to a so-called Bell-state measurement (BSM) after travelling through short standard telecommunication fibres. This measurement is performed by sending the two photons into the different input ports of a 50/50 beam splitter. Provided they exit through two different output ports and in different temporal modes (one photon early, the other one late), their joint state is projected onto the $\ket{\Psi^-}_{BC}=\frac{1}{\sqrt{2}}\left( \ket{e}_{B}\ket{\ell}_{C} - \ket{\ell}_{B}\ket{e}_{C}\right)$ Bell state, leaving the two 795~nm photons in the entangled $\ket{\Psi^+}_{AD}=\frac{1}{\sqrt{2}}\left( \ket{e}_{A}\ket{\ell}_{D} + \ket{\ell}_{A}\ket{e}_{D}\right)$ Bell state. We note that, in principle, it is also possible to make projection measurements onto the $\ket{\Psi^+}$ Bell state, thereby increasing the measurement's efficiency from maximally 25\% to 50\%  \cite{Valivarthi_14}. To detect the 1533~nm photons, we use free-running, tungsten silicide (WSi)-based superconducting nanowire single-photon detectors (SNSPD) cooled to around 0.9~Kelvin \cite{Marsili_13ab}. However, due to fibre transmission loss inside our cryostat, we find a system efficiency of around 50\%. Furthermore, we measure a detection-time jitter of 250~ps (FWHM), which is sufficiently small to allow resolving the  temporal qubit modes (spaced by 1.4 ns), and the dark-count rate of 10 Hz ensures little noise-pollution of detection signals. 

For the Bell-state measurement to work, the two 1532~nm photons must be indistinguishable at the beam splitter, i.e, their spatial, temporal, spectral, and polarization modes must be identical. This is verified using so-called Hong-Ou-Mandel (HOM) interference \cite{C. K. Hong 1987ab}: if two indistinguishable photons (not qubits) impinge on a symmetric beam splitter from different input ports, then they bunch and leave together by the same output port due to destructive interference between the probability amplitudes associated with both input photons being transmitted or both reflected. Conversely, if the two photons are distinguishable, no such interference occurs, and they leave the beamsplitter with 50\% probability through different outputs, resulting in coincident detections. The HOM visibility, defined as  $V_{HOM} = (N_{max}-N_{min})/N_{max}$ \cite{H. de Riedmatten 2003ab}, where $N_{max}$ and $N_{min}$ denote coincidence count rates measured with completely distinguishable and (maximally) indistinguishable photons, respectively, is a common way to characterize the degree of indistinguishability. We find $V_{HOM} = 27.5 \pm 2.5$\%. This value is consistent with the fact that our 1533~nm inputs are not single photons, but are mixtures of photon Fock states with thermal distribution, for which the HOM visibility is upper bounded by 1/3 \cite{H. de Riedmatten 2003ab}.

Repeating the measurement conditioned on the detection of  two 795~nm photons (described below) and with small pump power, which projects the 1533~nm inputs onto approximate single photons, the HOM visibility increases up to $87.5\pm 5.5$\%. We attribute the gap to the theoretical value of 100\% to insufficient spectral filtering of the photons, and, to a much smaller extent, to remaining contributions from the simultaneous emission of more than two photons. Indeed, using $V_{HOM}^{Max}= 1/ \sqrt{1+(\frac{\Delta T}{\tau})^2}$ \cite{M. Zukowski 1995ab, J. G. Rarity 1995ab}, where $\Delta T$ denotes the duration of the pump pulses, and $\tau$ the coherence time of the 1533~nm photons after filtering, we find that the maximum visibility achievable with our setup is 89\%, which corresponds to our result within experimental error.

To assess the quality of the entangled state after the entanglement swapping, the two 795~nm photons are sent through Mach-Zehnder interferometers that introduce the same travel-time difference of 1.4~ns as the interferometer that acts on the pump beam and thus allow projecting photon states onto various time-bin qubit states \cite{W.Tittel 2001ab}. To ensure phase stability during the measurements, i.e. constant projectors, the interferometers are passively temperature stabilized. Additionally, their phases are actively locked using a frequency-stabilized laser at 1550~nm wavelength that is also sent through the interferometers, and a home-made feedback loop. Finally, the 795~nm photons are  detected using four standard silicon avalanche photodiode-based single photon detectors with efficiencies around 50\%, detection jitter of 500 ps, and dark counts around a few hundred Hz. All detection signals are recorded using a time-to-digital converter that is started by a successful BSM. 

We also measure the total heralding efficiency~\cite{pomarico2012a}, which, for the 795~nm (1533~nm) photon, we define as the ratio between the rate of photon pair coincidence detections and the rate of the detection of 1533~nm (795~nm) photons. Hence, for the low mean photon number in our experiments, the heralding efficiency is equal to the probability of detecting a photon that has been created by downconversion in the source. This takes into account all transmission loss (such as loss in filter cavities, prisms and fibre coupling loss), non-unity detection efficiency, and a fundamental restriction caused by the bandwidth mismatch between the pump and downconverted photons. A simple calculation of the bandwidth restriction (see appendix) shows that the latter alone limits the heralding efficiency for the 795~nm (1533~nm) photons to 17.4\% (34.8\%). Considering furthermore the known values of 50\% (70\%) for detection efficiency, 40\% (80\%) for transmission through the cavity, and 85\% (85\%) for transmission through the prism, we anticipate a heralding efficiency of 2.96\% (16.6\%) for the 795~nm (1533~nm) photons.
Experimentally we obtain heralding efficiencies of 1.96\% for 795~nm and 5.8\% for 1533~nm photons. The difference between the measured and expected values is predominantly due to imperfect fibre coupling, which we assess to be 66\% and 35\% for the 795~nm and 1533~nm photons, respectively. Hence, we find that one of the main limitations to achieving higher heralding efficiencies is the bandwidth mismatch between the pump photon and the downconverted photons. This can be improved by using a spectrally narrower pump, as further described in the Appendix.

%

\section{Measurements and Results}

\begin{figure}[htpb]
\includegraphics[width=\columnwidth]{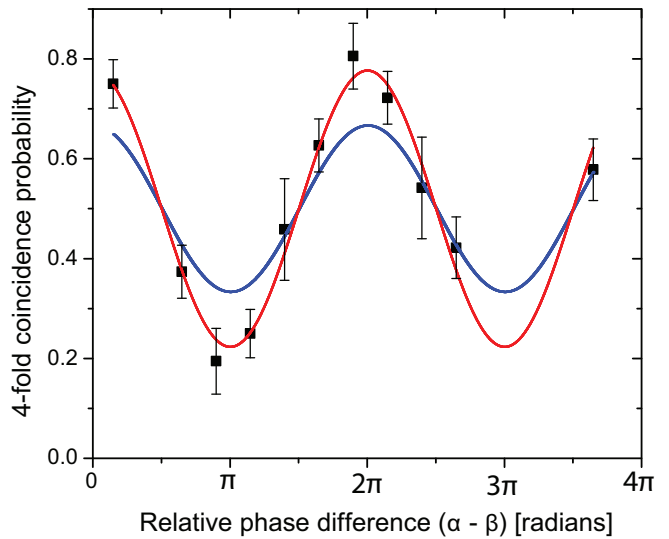}
\caption[Coincidence detection probabilities as a function of the phase difference in the measurement interferometers.]{color online) Coincidence detection probabilities (conditioned on a successful Bell-state measurement) of the 795~nm photons as a function of the phase difference in the measurement interferometers.  The uncertainty bars indicate one standard deviation and are calculated from measured rates assuming Poissonian detection statistics. The visibility of the sinusoidal fit is 56.2 $\pm$ 5.7\%. The blue curve features a visibility of 1/3, indicating the classical bound for a separable Werner state.}
\label{Probcoinc}
\end{figure}

To investigate the presence of entanglement between the two 795~nm photons after the BSM with the 1533~nm photons, first, we measure the probability for coincidence detection of two 795~nm photons in detectors $A$ and ${D}$, conditioned on projecting the two 1533~nm photons onto the $\ket{\Psi^-}$ Bell state, as a function of the phase difference $\alpha - \beta$ of the two analyzing interferometers. This corresponds to projecting each 795~nm photon onto equal superpositions of $\ket{e}$ and $\ket{\ell}$. (For these as well as the following measurements, the mean photon pair number was set to 19.1\%.) The results, depicted in Fig. \ref{Probcoinc}, show a sinusoidal curve with a fitted visibility of (56.2 $\pm$ 5.7)\%. The phase difference axis is recalibrated based on the fit so that $\alpha - \beta=0$ corresponds to a maximum of the coincidence probability. The sinus period is fixed to $2\pi$, however, refitting the data with the period as a free parameter yields an almost identical curve with a period of $(0.98\pm0.03) 2\pi$. This confirms the excellent stability and calibration of the relative interferometer phases during our measurements, which required a total of 36 hours (the average four-fold coincidence rate was about 10/hour). Furthermore, the visibility clearly exceeds the maximum visibility of 33\% that can be obtained using a separable Werner state  \cite{Perez96}. This confirms the presence of entanglement, provided the often made assumption of having a Werner state is satisfied.

\begin{figure}[htbp]
\begin{center}
\includegraphics[width=0.95\columnwidth]{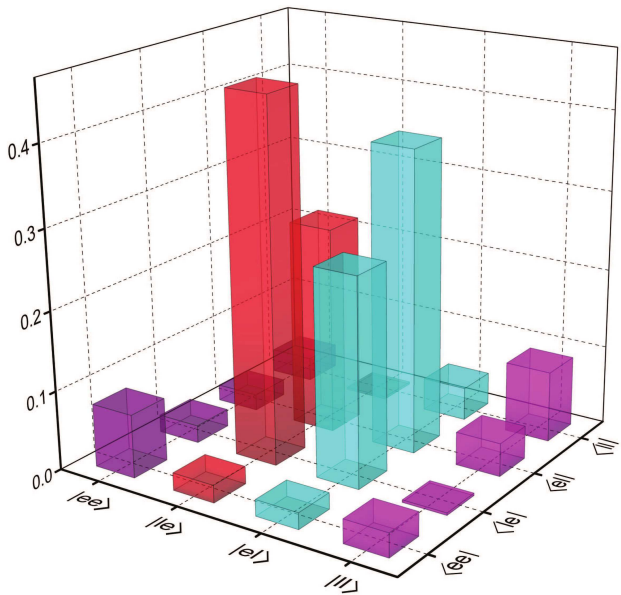}
\end{center}
\caption[Absolute values of the density matrix characterizing the joint state of the two 795~nm photons after entanglement swapping]{color online) Density matrix characterizing the joint state of the two 795~nm photons after entanglement swapping. The fidelity with respect to the $|\Psi^+\rangle$ Bell state -- which we find to be the nearest maximally entangled state -- is $(68 \pm 3)$\%.}
\label{entresult}
\end{figure}

To remove this assumption, we reconstruct the density matrix of the 795~nm photons' joint quantum state by means of maximum likelihood quantum state tomography (QST) \cite{Altepeter_05a}, which will allows assessing additional measures that quantify entanglement and derive more information about experimental imperfections. Towards this end we perform a total of 36 joint projection measurements corresponding to all combinations of projections onto eigenstates of $\sigma_X$, $\sigma_Y$ and $\sigma_Z$. Each combination requires a coincidence measurement (conditioned on a successful Bell-state measurement) with specific (local) interferometer phase settings (for projections onto eigenstates of $\sigma_X$ and $\sigma_Y$), or measurements of photon arrival times (for projections onto eigenstates of $\sigma_Z$). We emphasize the importance of interferometer phase stability, which we verified above, to ensure proper (and stable) measurements. The reconstructed density matrix is depicted in Fig. \ref{entresult}. It allows calculating the concurrence $(C)$, an entanglement measure that is zero for a separable state and larger than zero for an entangled state \cite{W. K. Wootters 1998ab}. We find $C=0.36 \pm 0.07$, which exceeds the threshold by five standard deviations (for this and the further experimental values the uncertainty is calculated by means of Monte Carlo simulation and assuming the coincidence counts follow Poissonian statistics), thereby confirming the conclusion of having an entangled state derived from the visibility in Fig.~\ref{Probcoinc}. We also compute our reconstructed state's fidelity with the expected state $\ket{\Psi^+}$ to be $F=(68\pm 3)\%$. Optimization over all maximally entangled states also shows that $\ket{\Psi^+}$ yields the largest fidelity with our reconstructed state (within the step size of the optimization algorithm).

Furthermore, we can now assess how well the reconstructed density matrix is described by a Werner state by calculating the fidelity of the swapped state with the nearest Werner state $\sigma=v \ket{\psi}\bra{\psi} +\frac{1}{4}(1-v) \mathbb{\opone} $. Here, $\ket{\psi}$ is any maximally entangled two-qubit state, $\mathbb{\opone}$ is the identity matrix and $v$ is a variable that parametrizes the state. Evaluating $F=(tr\sqrt{\sqrt{\sigma}\rho\sqrt{\sigma}})^2$ over all $\ket{\psi}$ and $v$, we find that the maximal value $ F=97.5^{+0.5}_{-4.4}\% $ is achieved for $\ket{\psi}=\ket{\Psi^+}$ and $v=0.592$.
Hence, using the fidelity measure indicates that the reconstructed state is close to a Werner state. 
The remaining difference is due to the admixture of non-white noise in the reconstructed state, which we primarily attribute to the residual distinguishability in the Bell-state measurement (this imperfection only affects measurements of $\sigma_x$ and $\sigma_y$ -- not $\sigma_z$ \cite{MDI-QKDmodel}). Hence, it is reasonable to use the fitted visibility in Fig.~\ref{Probcoinc}, which is based on measurements in the $\sigma_x$ and $\sigma_y$ bases, to infer the presence of entanglement in the state after entanglement swapping. 
However, the observation that our state is not an exact Werner state also leads to a note of caution, in that a visibility measurement comprising projections onto $\sigma_z$ -- which in our experiment is less affected by experimental imperfections  -- may provide an inflated measure of the entanglement in the swapped state.

To independently assess how experimental imperfections limit the amount of entanglement in the final bi-photon state, we employ the method described in \cite{S. Guha 2014a} with measured experimental parameters for heralding efficiency (1.96\% at 795~nm and 5.8\% at 1533~nm), HOM visibility (89\%), mean photon pair number (19\%) and fidelities (95\%) for the individual sources. This leads to a density matrix having a concurrence of 0.43, i.e. a state that is slightly closer to a maximally entangled state than the measured one. We find the fidelity of the density matrix predicted by this model and the one we measured to be $98.1^{+0.3}_{-4.0}\% $, which suggests that we understand the noise sources affecting the swapped state. Because this model accounts for non-white noise arising from the limited HOM visibility in the BSM, the predicted density matrix features a slightly larger fidelity with our measured state as that resulting from the comparison with the closest Werner state. 

Based on our analysis, we identify multi photon pair emissions, due to the probabilistic nature of SPDC sources, as our main source of errors that limits the value of the concurrence. To improve our result, we therefore have to decrease the probability of generating a photon pair per pulse below our current value of 19\%. For instance, reducing the mean photon pair emission probability to 10\% would, according to our model \cite{S. Guha 2014a}, result in an increase of the concurrence to 0.53 (keeping all other parameters unchanged). Note that alternative way to reduce the impact of multi-pair emissions are to exploit a quantum Zeno blockade to suppresses multi photons \cite{Y. -P. Huang 2012ab} or a quantum non-demolition measurement that reveals the number of simultaneously emitted photon pairs \cite{L. Liang 2014a}, thereby allowing in theory to completely ignore detections stemming from multi pair emissions. According to the model this would yield a concurrence of 0.82.


\section{Outlook and Conclusion}

Before concluding, let us briefly discuss the possibility to extend the current implementation into an elementary quantum repeater link, in which case the two 1533~nm photons have to travel tens of km before being submitted to the Bell-state measurement, and the two 795~nm photons have to be stored in optical quantum memories. We note that these criteria are easily met as, first, their long coherence time makes the 1533~nm photons robust against travel-time fluctuations during long-distance transmission, second, loss in optical fibres at this wavelength is minimal, and third, as quantum memories for 795~nm photons with 6 GHz bandwidth are available \cite{E. Saglamyurek 2011ab}. However, even assuming memory efficiencies exceeding 50\%, which remains to be demonstrated for memories of such large bandwidth, the coincidence count rates are currently too small to demonstrate an elementary quantum repeater link - let alone building a useful one. Solutions, on the one hand, are relaxed focussing of the 523~nm laser pulses into the SPDC crystals, which has been shown to significantly improve the coupling efficiency  \cite{Thew_13}. Secondly, the lossy cavities for the 795~nm photons can be removed as the memories themselves will work as filters. The latter will increase the coincidence rates by one order of magnitude. Furthermore, given the large bandwidth of the SPDC photons (of which we currently use only a few GHz), and the large bandwidth and spectral multi-mode  storage capacity of, e.g., atomic frequency comb-based Tm:LiNbO$_3$ quantum memories, it is possibility to work with many frequency channels in parallel using a quantum repeater architecture that we described in \cite{N. Sinclair 2014abc}.

To summarize, we have experimentally demonstrated the creation of entanglement between two 795~nm photons, whose properties are suitable for further storage in broadband quantum memories such as our Tm:LiNbO$_3$ waveguides, by means of a Bell-state measurement with two 1533~nm photons, each of which was initially entangled with one of the two 795~nm photons. Our demonstration constitutes an important step towards the generation of a quantum repeater: the heralded entanglement of two quantum memory-compatible photons. \\

\section*{Acknowledgements}
The authors thank Vladimir Kiselyov for support with electrical engineering. We gratefully acknowledge support through Alberta Innovates Technology Futures (AITF), the National Science and Engineering Research Council of Canada (NSERC), the US Defense Advanced Research Projects Agency (DARPA) Quiness and InPho Programs, and the Killam Trusts. Part of the research was carried out at the Jet Propulsion Laboratory, California Institute of Technology, under a contract with the National Aeronautics and Space Administration. W.T. is a senior fellow of the Canadian Institute for Advanced Research (CIFAR).

\section*{Appendix: Heralding efficiency vis-a-vis photon bandwidth}

The bandwidth mismatch between the pump light and the filters used in our experiment for the downconverted photons fundamentally limits the maximum attainable heralding efficiency for each of the down-converted photons (commonly referred as signal and idler). In this section we develop a simple model -- inspired by the pictorial representation in~\cite{palmett2013a} -- that captures the main consequences of this effect.

As described in the main text, we define the heralding efficiency for the signal ($\eta_{H_{s}}$) as the ratio between the coincidence detection rate ($C_{si}$) and the single detection rate for the idler ($S_{i}$):
\begin{equation}
\eta_{H_{s}}=\frac{C_{si}}{S_{i}}.
\end{equation}
Similarly, the idler heralding efficiency is found as:
\begin{equation}
\eta_{H_{i}}=\frac{C_{si}}{S_{s}}.
\end{equation}

\begin{figure}[htbp]
\begin{center}
\includegraphics[width=1\columnwidth]{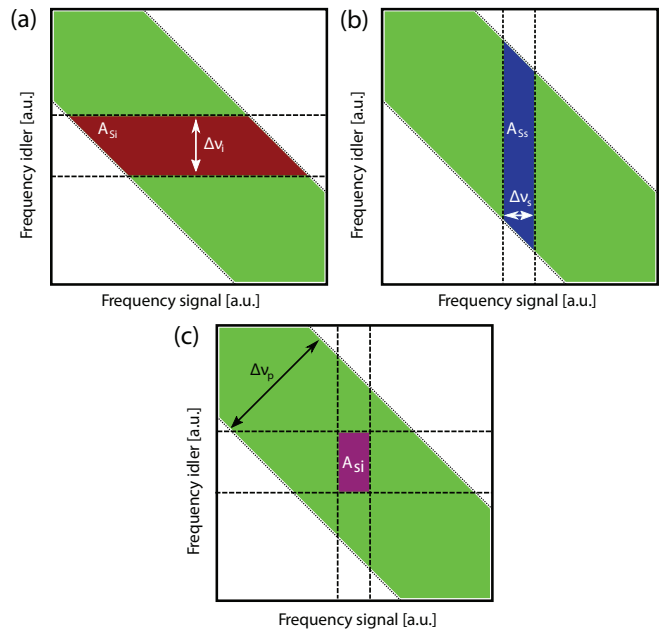}
\end{center}
\caption{color online) Joint Spectral Amplitude of photon pairs generated via an SPDC process with bandwidths of pump-pulse and spectral filters indicated.}
\label{heralding}
\end{figure}

The pump laser has a spectral bandwidth $\Delta\nu_p$ given by the pump pulse duration. Since the down-conversion process must conserve energy, the frequencies of any pair of down-converted photons are correlated to within the pump bandwidth. As a result, the Joint Spectral Amplitude (JSA) of the down-converted photons (see Fig. \ref{heralding}) can be illustrated by a diagonal band (green region) with a cross sectional width corresponding to the spectral width of the pump ($\Delta\nu_p$). Note that we assume that, over the relevant bandwidth, the phase-matching condition is less restrictive than that resulting from energy conservation, and thus we do not indicate it on the figure. However, phase-matching determines the total extent of the diagonal band (outside the view of the figure) and thus sets the overall bandwidth of down-converted photons. 

Next we explore the effect of restricting the bandwidth of either the signal or the idler photon by means of spectral filters with bandwidths $\Delta \nu_s$ and $\Delta \nu_i$, respectively. In Fig.~\ref{heralding}, and for the case in which we filter the idler, this corresponds to carving out a horizontal band in the JSA. The idler photon single detection rate is proportional to the overlap of this horizontal band with the diagonal band representing energy conservation (red area in Fig.~\ref{heralding}a), i.e. $A_i=\sqrt{2} \Delta\nu_p \Delta\nu_i$. Similarly, for the case of filtering the signal, the single count rate is proportional to the area of the blue region in Fig.~\ref{heralding}b, which is $A_s=\sqrt{2} \Delta\nu_p \Delta\nu_s$. 
These expressions are approximate because they assume that the count rates of the filtered photons are directly proportional to the specified areas. This, however, is only correct if the spectral profile of the filters and the JSA given by the pump bandwidth are box-shaped (flat-top). Moreover, the area is only calculated as above if the filter bandwidths are smaller than the pump-bandwidth i.e. $\Delta\nu_{s(i)} < \Delta\nu_p$.

Finally, we extend this picture to the case in which we filter both signal and idler and measure the coincidence count rate. As illustrated by Fig.~\ref{heralding}c the coincidence count rate is proportional to the area (in pink), i.e. the intersection of the horizontal and vertical bands given by the filters -- $A_{si}=\Delta\nu_s \Delta\nu_i$. Clearly, this area is smaller than both $A_s$ and $A_i$, thus limiting the heralding efficiency. More precisely the heralding efficiencies can be expressed as:
\begin{align}\label{heraldeff}
\eta_{H_{s}}&=\frac{C_{si}}{S_{i}}=\frac{A_{si}}{A_{i}}=\frac{\Delta \nu_s}{\sqrt{2} \Delta \nu_p} \nonumber \\
\eta_{H_{i}}&=\frac{C_{si}}{S_{s}}=\frac{A_{si}}{A_{s}}=\frac{\Delta \nu_i}{\sqrt{2} \Delta \nu_p}.
\end{align}

In our experiments $\Delta \nu_s=6$~GHz (795 nm photon), $\Delta \nu_s=12$~GHz (1533~nm photon) and $\Delta \nu_p=24.4$~GHz (523~nm photon). Using these values in Eq.~\ref{heraldeff}, we obtain $\eta_{H_{s}}=17.4\%$ and $\eta_{H_{i}}=34.8\%$. Though these values are calculated using a simple model, they provide a qualitative explanation of the maximum attainable heralding efficiency for each of the signal and idler photons in our system and allows us to assess the loss due to other factors such as optical elements, optical fibre coupling, and detector efficiency.

\bibliographystyle{apsrev4-1}

\end{document}